# IETF protocol suite for the Internet of Things: Overview and Recent Advancements

Roberto Morabito*[†], Jaime Jiménez*
*Ericsson Research, Finland
[†]Princeton University, USA
{roberto.morabito, jaime.jimenez}@ericsson.com

*Abstract*—Internet of Things (IoT) is a rapidly growing technological domain and the effort that many Standards Developing Organizations (SDOs) and alliances are dedicating to it is constantly increasing. The Internet Engineering Task Force (IETF) is certainly one of the most active SDOs operating in this broad domain. With the aim of building a comprehensive, interoperable and streamlined IoT stack, several IETF working groups are developing protocols, in multiple technological domains, which are already nowadays very relevant to the IoT. This article gives a concise but comprehensive survey of the IETF efforts in the IoT. We discuss mainstream standardization and research activities, as well as other IETF-connected supporting initiatives provided still in the context of IoT.

## I. Introduction

In the last two decades, the Internet of Things (IoT) has become a significant research and development opportunity for industry, academia, and SDOs [1]. Simplifying its definition in non-rigorous terms, IoT refers to the usage of Internet protocols for enabling communication from "things" toward humans, between "things" and more complex Internet-based infrastructures. A key aspect of the IoT is the possibility of using Internet protocol technologies on constrained devices such as sensors and actuators. These are devices characterized by very limited computational capabilities, with a supported code complexity ranging between 100 and 250 kB, RAM between 10 and 50 kB, and limited resources also in battery, bandwidth, and connectivity [2].

However, the integration of the the well-known TCP/IP Internet protocol suite [3] on top of constrained devices generates several challenges, since such protocols were not originally designed to operate with this type of nodes. For example, the design guidelines followed for such protocols – originally defined for the Web – may not always be suitable for constrained network environments, in which nodes are limited in CPU, memory, and the networks are often characterized by high packet loss, low throughput, frequent topology changes and small useful payload sizes [4].

The Internet Engineering Task Force (IETF) has long been on the front line in overcoming such challenges, by leading the standardization of Internet communication protocols suitable for resource-constrained devices and networks. Where possible, IETF aims to make existing Internet protocols suitable for IoT scenarios, by providing minor tweaking on their original characterization and by taking into account the different requirements. In other cases, significant gains can only be achieved by defining new protocols.

The authors would like to thank the Ericsson colleagues, Carsten Bormann, and Emmanuel Baccelli who provided helpful feedback and insightful comments on the paper.

Another key guideline followed by IETF in IoT-oriented standardization activities is the possibility of enabling a wide range of "things" to use interoperable technologies for communicating with each other – in this regard, "things" can range from embedded sensors to complex machinery (e.g., a car) or even large infrastructure (e.g. a bridge). As evidence of this demanded IoT interoperability, Figure 1 shows how an IoT stack deployable nowadays on constrained devices can include several IETF standards (defined for the IoT context and beyond), as well as non-IETF standards.

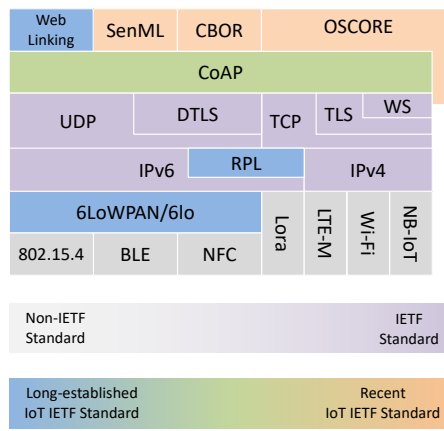

Fig. 1: An example of IoT device stack comprising a mix of IETF and Non-IETF standards.

The depicted IoT stack represents just one possible example of an IoT device stack. It could easily comprise many additional protocols, including ones that were not originally designed for IoT or in IETF, which are now widely used in this context – e.g., the ISO standard Message Queue Telemetry Transport (MQTT) or the IETF standard HTTP. This is mainly caused by the fact that IoT lends itself to a significant and varied number of definitions, design visions, and deployments [5].

However, since we focus on IETF and its activities in the constrained IoT area, we rely on the vision and definition that IETF itself provides about IoT in [6]:

*"The Internet of Things (IoT) refers to devices, that are often constrained in communication and computation capabilities, now becoming more commonly connected to the Internet, and to various services that are built on top of the capabilities these devices jointly provide. It is expected that this development will usher in more machine-to-machine communication using the Internet with no human user actively involved"*.

IETF activities are also supplemented by the Internet Re-search Task Force (IRTF). While the first one focuses on the shorter-term matters of Internet engineering and standards making, the second one works on longer-term subjects with a clear research-oriented approach.

Figure 2 illustrates different Internet-related technological meta-domains, whose development is extensively enhanced by IETF and IRTF. In the map, we identify four main areas (i.e. *Connectivity*, *Routing*, *Application*, *Security*) which can be mapped a substantial number of IETF working groups characterized by a clear IoT scope. Additionally, several other working and research groups can somehow be classified as IoT-related, although characterized by a different general purpose. We map these additional working groups into two specific categories – *Experimental & Use Cases* and *Infrastructures*.

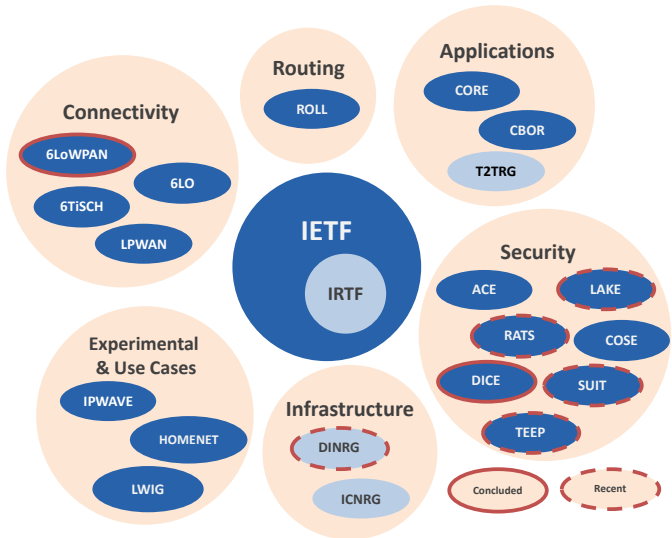

Fig. 2: IETF Working groups operating in the IoT classified in meta-domains.

The remainder of this article focuses on analyzing, at our own discretion, the leading IETF and IRTF activities currently oriented toward IoT in the technical domains introduced above. We focus, in particular, on the progress of the last two years. The article is organized as follows. We first focus on the connectivity and routing aspects. Afterward, application and security areas are reviewed and discussed, respectively. Following, we overview ongoing standardization efforts related to experimental activities, use cases, and infrastructures. Before the conclusions are given, we also introduce the other committees and organizations established to work alongside IETF and their areas of responsibility in the IoT context. This article is intended to provide an update of [7], by providing an up to date overview on the consolidated and emerging activities on IoT pursued within IETF.

**Terminology**. Before moving forward in our analysis, it is worth clarifying the terminology commonly used in such standardization context. The following description can help the reader become familiarized with specific terms and acronyms, which frequently recur in most of the IETF processes and in this article.

- *IETF Working Group (WG)* is the primary mechanism for the development of IETF specifications and guidelines, many of which are intended to be standards or recommendations.
- *IRTF Research Group (RG)* explores and works on research-related topics with a longer-term approach when compared to a WG. However, its organizational nature remains similar.
- *Internet-Drafts (I-Ds)* are working documents of the IETF, its areas, and its Working Groups. Note that other groups may also distribute working documents as I-Ds. I-Ds can be proposed by individual parties, normally within the scope a specific WG. In order to become WG I-Ds, they need to go through an adoption process within the WG itself.
- *Request for Comment (RFC)* is a technical and organizational publication, which describes mechanisms, implementations guidelines, or innovations applicable on topics related to Internet protocols, Internet-connected systems, applications, architecture and technology. RFCs can follow multiple tracks; standard, informational or experimental. Standard track RFCs that reach an adequate level of maturity become firstly *proposed standard*. Once a proposed standard reaches a high degree of maturity and adoption, it can become *Internet standard*. In order to become an Internet standard, it is expected that multiple interoperable implementations must exist (which are often released as open source).

## II. CONNECTIVITY & ROUTING

IETF's work on IoT started by bringing connectivity to the constrained space, already at the turning of this century. At that time, there were already different Wireless Sensor Network (WSN) proposals that were not making use of standard IETF network protocols. However, the work for connecting constrained devices to the Internet by enabling IPv6 addressing on them has represented one of the first step towards the IoT.

### A. Connectivity

**6LoWPAN** — The IPv6 over Low power WPAN (6LoWPAN) WG operated from March 2005 to January 2014. The group produced several RFCs covering application scenarios and use cases for Low-power Wireless Personal Area Networks (LoWPANs). At that time, LoWPANs were usually characterized by proprietary networking mechanisms. Thus, the group purpose was to bring IP (IPv6 in particular) to short range, low bit rate, low power radios – specifically, IEEE 802.15.4 [8]. Along with header compression, one of the big hurdles existing for enabling such integration was due to the mismatch between the maximum frame size of 127 bytes and the Maximum Transmission Unit (MTU) of IPv6. This has led to the specification of an adaptation layer that provides fragmentation and reassembly, and allows the above mentioned problem to be overcome. Table I shows three among the most representative RFCs produced in 6LoWPAN.

**6TiSCH** — The WG started in 2013 and still continues. It focuses on enabling IPv6 over the Time-Slotted Channel Hopping (TSCH) MAC mode of the IEEE802.15.4 standard.

The TSCH WG brings to IoT the possibility of time synchronization for IoT devices, which results in less battery consumption. Its activities particularly target industrial scenarios, where demanding network conditions may occur and highly reliable delivery is needed. As shown in Table I, the WG has so far focused on different aspects ranging from defining guidelines for 6TiSCH-related use cases and scenarios, to the definition of minimal IP over TSCH configuration required on a device, as well as contextualization and protocol specification for the IPv6 over the TSCH mode of IEEE 802.15.4e network – known as 6TiSCH. The group is also defining the minimal requirements for a device to join the aforementioned 6TiSCH network securely. In this regard, it has prioritized the use of two CoRE WGs' specification: Constrained Application Protocol (CoAP) for the messaging and Object Security (OSCORE) to secure the communication.

**6lo** — Initiated in 2013, the IPv6 over Networks of Resource-constrained Nodes (6lo) WG has focused – like 6LoWPAN did – on bringing IPv6 connectivity over constrained node networks of different types. While 6LoWPAN specifically targeted 802.15.4, 6lo opened that focus up to other network technologies, putting together all the efforts made to compress protocol headers and extension headers. 6lo has produced guidelines for IPv6 over various constrained networks such as Bluetooth Low Energy (BLE), ITU-T G.9959 networks, Digital Enhanced Cordless Telecommunications (DECT) Ultra Low Energy (ULE) or even over Master-Slave/Token-Passing (MS/TP) wired networks. The WG has also produced specifications on information and data models (e.g., MIBs for managing 6LoWPANs), adaptation layer extensions (e.g., Generic Header Compression for IPv6 over 6LoWPAN), as well as other maintenance documents such as new IEEE Ethertype assignments, RFC updates or informational RFCs on privacy considerations. The group is still active, working on bringing IPv6 to other networks such as near field communication (NFC).

**LPWAN** — The WG focuses on enabling IPv6 connectivity over multiple Low Power Wide-Area Network (LPWAN). LPWANs devices possess large coverage at the cost of substantially lower bandwidth (e.g., order of hundreds of millibits per second [9]) and reduced duty cycle. The group has selected four LPWANs to work on: Long Range (LoRa), Wireless Smart Utility Networks (WI-SUN), Narrowband IoT (NB-IoT), and Sigfox. The WG started its work in 2016 and has completed an overview document specification of their selected LPWAN technologies. LPWAN's current main focus is on LPWAN static context header compression (SCHC) for CoAP and for UDP (Table I).

*B. Routing*

**ROLL** — The Routing Over Low-power and Lossy networks (ROLL) WG focuses on routing solutions, targeting specific deployment environments such as connected home, building, and urban sensor networks. The scope of ROLL has been re-defined several times. The main outcome of its earlier activities is the specifications of the IPv6 Routing Protocol for Low-Power and Lossy Networks (RPL) protocol (RFC 6550, described in Table I), an the IPv6 Routing Protocol for Low-Power and Lossy Networks (LLNs) that takes into consideration various aspects including high reliability and connectivity while ensuring low resource utilization into the devices. More recently, the WG has kept maintaining and improving already developed protocols (e.g., RPL), as well as defining related extensions for routing metrics, objective functions, and multicast. Finally, the WG has also produced several documents concerning requirements and applicability statements for routing in LLNs, terminology, and security threat analysis.

TABLE I: Selection of relevant RFCs and I-Ds in the connectivity and routing domain.

| RFC number and Title | Description |
|---|---|
| **Connectivity** | |
| *6lowpan* | |
| RFC 4944 — Transmission of IPv6 Packets over IEEE 802.15.4 Networks | Describes the frame format requirements for the transmission of IPv6 packets on IEEE 802.15.4 networks. Additional aspects discussed are definition for forming IPv6 link-local addresses, provisions for packet delivery in IEEE 802.15.4 meshes, IPv6 header compression, etc. |
| RFC 6282 — Compression Format for IPv6 Datagrams over IEEE 802.15.4-Based Networks | Updates RFC 4944. Specifies an IPv6 header compression format for IPv6 packet delivery in 6LoWPANs. It also defines a compression mechanism for multicast addresses, as well as UDP header compression. |
| RFC 6775 — Neighbor Discovery Optimization for IPv6 over Low-Power Wireless Personal Area Networks (6LoWPANs) | This document describes simple optimizations to IPv6 Neighbor Discovery, its addressing mechanisms, and duplicate address detection for Low-power Wireless Personal Area Networks and similar networks. |
| *6tisch* | |
| RFC 7554 — Using IEEE 802.15.4e Time-Slotted Channel Hopping (TSCH) in the Internet of Things (IoT): Problem Statement | Describes the environment, problem statement, and goals for using the Time-Slotted Channel Hopping (TSCH) Medium Access Control (MAC) protocol of IEEE 802.15.4e in the context of Low-Power and Lossy Networks (LLNs). |
| RFC 8480 — 6TiSCH Operation Sublayer (6top) Protocol (6P) | Describes the "IPv6 over the TSCH mode of IEEE 802.15.4e" (6TiSCH) Operation Sublayer (6top) Protocol (6P), which enables distributed scheduling in 6TiSCH networks. |
| *6lo* | |
| RFC 7428 — Transmission of IPv6 Packets over ITU-T G.9959 Networks | Describes the frame format for transmission of IPv6 packets as well as a method of forming IPv6 link-local addresses and statelessly autoconfigured IPv6 addresses on ITU-T G.9959 networks. |
| RFC 7668 — IPv6 over BLUETOOTH(R) Low Energy | Describes how low-power Bluetooth standard (version 4.0 onwards) transports IPv6 exploiting LoWPAN techniques (6LoWPAN). |
| RFC 8105 — Transmission of IPv6 Packets over Digital Enhanced Cordless Telecommunications (DECT) Ultra Low Energy (ULE) | Describes how IPv6 is transported over DECT ULE using IPv6 over Low-Power Wireless Personal Area Network (6LoWPAN) techniques. |
| RFC 8163 — Transmission of IPv6 over Master-Slave/Token-Passing (MS/TP) Networks | This specification defines the frame format for transmission of IPv6 packets and the method of forming link-local and statelessly autoconfigured IPv6 addresses on MS/TP networks. |
| *lpwan* | |
| RFC 8376 — Low-Power Wide Area Network (LPWAN) Overview | Provides an informational overview of the key LPWAN technologies, which are highly considered by IETF, and analyzes their gaps for fully enable IP communication in LPWANs. |
| Draft Title | Description |
| *6tisch* | |
| Minimal Security Framework for 6TiSCH | Describes the steps required for a new device to securely join a 6TiSCH network. The framework involves the use of a central entity that enables the sharing of a symmetric key with the device. The framework defines the Constrained Join Protocol and its CBOR data structure, as well as configures the entire 6TiSCH communication stack for securing the device joining. |
| An Architecture for IPv6 over the TSCH mode of IEEE 802.15.4 | Defines the required components of a network architecture that meets the requirements of Low-Power wireless deterministic applications (e.g., low-latency, low-jitter and high-reliability packet delivery). In order to satisfy such requirements, it combines high-speed powered backbone and sub-networks using IEEE 802.15.4 TSCH. |
| *6lo* | |
| Transmission of IPv6 Packets over Near Field Communication | This document describes how IPv6 is transmitted over NFC using 6LowPAN techniques. |
| *lpwan* | |
| Static Context Header Compression (SCHC) and fragmentation for LPWAN, application to UDP/IPv6 | Defines a framework designed for LPWAN called Static Context Header Compression (SCHC), which enables both header compression and fragmentation. It defines a header compression mechanism and its application to compress IPv6/UDP headers, as well as fragmentation and reassembly mechanism that are used to support the IPv6 MTU requirement over the LPWAN technologies. |
| LPWAN Static Context Header Compression (SCHC) for CoAP | Defines how to apply SCHC header compression in CoAP headers, accounting for the different CoAP (flexible) header structure when compared with IPv6 and UDP protocols. |
| SCHC over NB-IoT | Describes a profile of use of SCHC over the NB-IoT wireless access and provides elements for an efficient parameterization. |
| Static Context Header Compression (SCHC) over LoRaWAN | Describes a profile of use of SCHC over LoRaWAN networks and provides elements for an efficient parameterization. |
| **Routing** | |
| RFC number and Title | Description |
| *roll* | |
| RFC 6550 — RPL: IPv6 Routing Protocol for Low-Power and Lossy Networks | Specifies an IPv6 Routing Protocol (RPL) for Low-Power and Lossy Networks (LLN), which are networks whose characterizing components (e.g. routers and interconnected devices) are constrained. |

### III. APPLICATION

In the application domain, there are currently three groups targeting the IoT context.

Constrained RESTful Environments (CoRE) is one of the key WGs, also in view of its well-established activities started already in 2010. More recently, two additional groups have been chartered in the IoT application domain, namely the Thing-to-Thing (T2TRG) RG in 2015 and the Concise Binary Object Representation Maintenance and Extensions (CBOR) WG in 2017.

**CoRE** — The main goal of CoRE is to define mechanisms and frameworks that are suitable for resource-oriented applications, intended to run on constrained IP networks (i.e., networks characterized by operating constraints like those defined in ROLL and 6Lo). CoAP, defined in RFC 7252, represents the foundation around which most of CoRE activities are developed [10]. CoAP is a UDP-based protocol that enables manipulation of resources hosted by devices and it is considered analogous to HTTP for constrained networks (although CoAP is designed for being used also in non-constrained IP networks). Based on CoAP specification, CoRE explores several additional areas (e.g., security, interoperability, data sensors representation, etc.), with the purpose of building a comprehensive and standard-compliant ecosystem around it. Table II summarizes some among the most relevant RFCs and I-Ds produced by the WG. Although CoAP was originally designed in order to operate over UDP and Datagram Transport Layer Security (DTLS), CoRE is committing to ensure a functional mapping of CoAP towards additional transport protocols and security mechanisms (e.g., TCP and TLS in RFC 8323), application protocols (e.g., HTTP in RFC 8075). Interoperability with non-IP networks and non-IETF standards and implementations also represents a key requirement. CoRE standards are used by other SDOs like OMA SpecWorks [11] which uses CoAP on its Lightweight Macchine-to-Machine (LWM2M) IoT device management protocol.

TABLE II: Selection of relevant RFCs and I-Ds in the application domain.

| RFC number and Title | Brief summary |
|---|---|
| core | |
| RFC 6690<br>Constrained RESTful Environments (CoRE) Link Format | Defines a specific link format for being used by constrained web servers. The Constrained RESTful Environments (CoRE) Link Format is carried as a payload and allows describing hosted resources, their attributes, and other relationships between links. |
| RFC 7252<br>The Constrained Application Protocol (CoAP) | Defines the Constrained Application Protocol (CoAP), which is a RESTful application protocol mainly designed for being used in machine-to-machine (M2M) application, constrained nodes and constrained networks. Among the different CoAP features described in this document, we can find definition of the request/response interaction model between application endpoints, support for built-in discovery of services and resources, and definition of Web-related key concepts such as URIs and Internet media types. |
| RFC 7641<br>Observing Resources in the Constrained Application Protocol (CoAP) | Defines a CoAP extension for observing the resource state of a CoAP server. Specifically, it enables CoAP clients to be kept updated about the representation of a resource over a period of time. |
| RFC 8323<br>CoAP (Constrained Application Protocol) over TCP, TLS, and WebSockets | Delineates the changes required to use CoAP over TCP, TLS, and WebSockets transports. |
| RFC 8428<br>Sensor Measurement Lists (SenML) | Defines a format for representing simple sensor measurements and device parameters in Sensor Measurement Lists (SenML). It offers several representational notations, JSON, CBOR, EXI and XML. |
| RFC 8613<br>Object Security for Constrained RESTful Environments (OSCORE) | Defines a method for application-layer protection of the Constrained Application Protocol (CoAP), using CBOR Object Signing and Encryption (COSE). OSCORE is specifically designed for constrained environments and provides end-to-end protection between endpoints communicating using CoAP (or CoAP-mappable HTTP). |
| cbor | |
| RFC 8610<br>Concise data definition language (CDDL): a notational convention to express CBOR and JSON data structures | Defines a notational convention to express in an easy and unambiguous way structures for protocol messages and data formats that use CBOR (RFC 7049) or JSON. |
| t2trg | |
| RFC 8576<br>Internet of Things (IoT) Security: State of the Art and Challenges | Discusses the lifecycle stages of a thing, followed by an analysis of security threats to whom the thing may be subjected and the challenges to deal with in order to secure against these possible threats. The draft concludes with a series of recommendations to follow for the deployment of secure IoT systems. |
| Draft Title | Description |
| core | |
| CoRE Resource Directory | Defines an entity called Resource Directory (RD), which tries to solve the problem of direct discovery of resources hosting registrations of resources held on other servers and allowing lookups to be performed for those resources. |
| The Constrained RESTful Application Language (CoRAL) | Defines the Constrained RESTful Application Language (CoRAL), which features a data and interaction model, but also two specialized serialization formats for the description of typed connections between resources on the Web ("links"), possible operations on such resources ("forms"), as well as simple resource metadata. |
| CoAP Management Interface | Describes CoMI (CoAP Management Interface), a network management interface that uses CoAP methods to access datastore and data node resources defined in YANG. CoMI's use of YANG to CBOR mapping mechanisms and YANG Schema Item iDentifier (SID) is defined as CORECONF. |
| cbor | |
| Concise Binary Object Representation (CBOR) | Defines the Concise Binary Object Representation (CBOR) data format, which is designed to satisfy the requirements of extremely small code size, fairly small message size, and extensibility. |
| t2trg | |
| RESTful Design for Internet of Things Systems | Provides a detailed guidance for designing IoT systems that fully rely on the Representational State Transfer (REST) architectural style. |

**CBOR** — The WG has as its primary objective the maintenance of RFC 7049, which defines the Concise Binary Object Representation (CBOR) data format. Originally, CBOR was defined to make available the benefits that JavaScript Object Notation (JSON, RFC 7159) brings to the web towards the constrained devices, and today there is available a CBOR library in almost any programming language. CBOR includes binary data as well as an extensibility model using a binary representation format that is easy to generate and parse.

However, in view of its increasing usage as a message format in other WGs (e.g., CORE, ANIMA), the group is aiming to revise its definition by taking into account the related security matters. The new CBOR definition will ensure backward compatibility, and it will be designed in such a way that a specific data definition language (CDDL; Table II) provides a single description format to express CBOR and JSON encoded messages.

**T2TRG** — The WG investigates open research questions stemming from the use of well-established standards already defined in the scope of IoT-oriented IETF WGs. It additionally explores emerging opportunities and issues arising from new requirements of IoT infrastructures, which consequently exhibit standardization potential in IETF. Although in this context the term "thing" often refers to constrained nodes, this RG targets IoT architectures where not all things are highly constrained. Leaving aside the nodes' computational capabilities, it is assumed that devices can communicate among themselves and with the rest of the Internet. T2TRG is a research-oriented group, in which the topics explored explored vary considerably, ranging from the management and operation of constrained-node networks, security and lifecycle management, methods to use the REST paradigm in IoT scenarios, and semantic interoperability – to this extent, although we categorized this WG in the application meta-domain, a relevant part of the WG's activities belong also to other domains. Table II describes a very limited part of the considerable documentation produced within the T2TRG. Another relevant aspect characterizing T2TRG is interaction with other standardization bodies and consortia active in the IoT area. For example, with the W3C Web of Things (WoT) interest group [12], how web technologies can be harnessed in the future of IoT is explored.

## IV. SECURITY

**DICE** — Initiated in 2013, the DTLS In Constrained Environments (DICE) WG completed its work in 2016. DICE focused on defining guidelines and mechanisms for the support of DTLS (RFC 6347) in the context of constrained environments (including constrained devices and constrained networks). Essentially, DICE provided a dedicated DTLS profile suitable for IoT applications (RFC 7925, Table III), where CoAP is assumed to be the main protocol used to manage such resources.

**ACE** — The main objective of the WG is the definition of authentication and authorization mechanisms suitable for resource access in constrained environments. Three-party authentication and authorization protocols previously defined in IETF (e.g., Public Key Infrastructure – PKI, Web Authorization Protocol – OAuth) are mostly suitable for non-constrained environments and do not consider additional requirements and limitations of typical IoT scenarios; for example, intermittent connectivity may reduce the possibility to contact an authorization server in real-time. Assuming that CoAP over DTLS is used to access resource-constrained servers, ACE seeks to assess the suitability of already existing protocols as baseline for the definition of its analogue in constrained environments. Table III describes only a limited subset of the WG outcomes, allowing one to understand which kind of contribution is produced by this WG. At the moment an OAuth-based framework with profiles for DTLS and OSCORE is nearing completion, while several related individual-submission I-Ds that address group communication are in the process of WG adoption.

TABLE III: Selection of relevant RFCs and I-Ds in the security domain.

| RFC number and Title | Description |
|---|---|
| **ace** | |
| RFC 7744 *Use Cases for Authentication and Authorization in Constrained Environments* | Presents authentication and authorization mechanisms for the entire life cycle of a constrained device. |
| **dice** | |
| RFC 7925 *Transport Layer Security (TLS) / Datagram Transport Layer Security (DTLS) Profiles for the Internet of Things* | Profiles two widely deployed Internet security protocols for IoT systems: Transport Layer Security (TLS) and Datagram Transport Layer Security (DTLS) 1.2. |
| **cose** | |
| RFC 8152 *CBOR Object Signing and Encryption (COSE)* | Defines the CBOR Object Signing and Encryption (COSE) security mechanisms for the CBOR data format such as creation and processing of signatures, message authentication codes, representation of cryptographic keys using CBOR, etc. There is currently an ongoing attempt to update this specification that would make RFC 8151 obsolete. |

| Draft Title | Description |
|---|---|
| **ace** | |
| *Authentication and Authorization for Constrained Environments (ACE) using the OAuth 2.0 Framework (ACE-OAuth)* | Defines ACE-OAuth, which is an authentication and authorization framework specially designed for constrained networks and IoT devices. The mechanism employs several building blocks including OAuth 2.0 and CoAP. |
| *Datagram Transport Layer Security (DTLS) Profile for Authentication and Authorization for Constrained Environments (ACE)* | Defines a profile for the ACE framework that uses DTLS for communication security between entities in a constrained network using either raw public keys or pre-shared keys. |
| *OSCORE profile of the Authentication and Authorization for Constrained Environments Framework* | Defines a profile for the ACE framework that uses OSCORE to secure a CoAP communication. |
| **cose** | |
| *CBOR Object Signing and Encryption (COSE): Initial Algorithms* | Defines an initial set of algorithms that are used for COSE signing and encryption. |
| **suit** | |
| *A Firmware Update Architecture for Internet of Things* | Defines requirements and architecture for a firmware update mechanism suitable for constrained IoT devices. The architecture is defined in such a way to decouple it from the underlying transport of the firmware images and associated meta-data. |
| *An Information Model for Firmware Updates in IoT Devices* | Defines all the information that must be present in the meta-data ("manifest") used for ensuring a secure firmware update mechanism. The manifest describes the firmware image(s) and ensures appropriate protection. |
| *A Concise Binary Object Representation (CBOR)-based Serialization Format for the Software Updates for Internet of Things (SUIT) Manifest* | Describes a serialization format in CBOR of a manifest about the firmware for an IoT device (e.g where to find the firmware, the devices to which it applies, and cryptographic information protecting the manifest). |
| **teep** | |
| *Trusted Execution Environment Provisioning (TEEP) Architecture* | Defines architecture components of a protocol for handling the lifecycle of applications running inside a Trusted Execution Environment (TEE). |
| *HTTP Transport for Trusted Execution Environment Provisioning: Agent-to-TAM Communication* | The Trusted Execution Environment Provisioning (TEEP) Protocol is used to manage code and configuration data in a Trusted Execution Environment (TEE). |
| **rats** | |
| *The Entity Attestation Token (EAT)* | An Entity Attestation Token (EAT) provides a signed (attested) set of claims that describe state and characteristics of an entity, typically a device like a phone or an IoT device. |

**SUIT** — The Software Updates for Internet of Things (SUIT) WG was chartered at the end of 2017 with the goal to strengthen, from security perspective, software update mechanisms so as to mitigate potential attacks (e.g., Mirai DDOS attack in 2017). Specifically, the group aims to define a firmware update solution to be used even with Class 1 type of devices [2], i.e., devices with 10 KiB RAM and 100 KiB flash. The design of a manifest format that provides meta-data about the firmware image is currently under development. The manifest also specifies pre- and post- conditions for the installation of the firmware as well as other parameters that need to be validated (e.g., hardware conditions, processor information, firmware makers, etc.).

**TEEP** — The Trusted Execution Environment Protocol (TEEP) WG was chartered in 2018. The WG operates in the context of a specific and highly secure area of modern devices' processor, as well as all the Trusted Application (TA) components executed in it. Currently two drafts are being developed in the TEEP WG that provide an overview of the architecture and protocol specifics. Its architecture document presents the need for an interoperable protocol for managing TAs that run in different TEEs of various devices. The document also contains definitions of the parties involved in the protocol and the requirements when it comes to TEE attestation. The group may document several attestation technologies considering different hardware capabilities, performance, privacy, and operational properties.

**COSE** — The CBOR Object Signing and Encryption (COSE) WG was chartered in 2015, concluded in 2016, and re-chartered in 2019. The group makes use of CBOR for object signing and encryption formats. One motivation is to reuse the cryptographic keys, message authentication (MACs), encryption, and digital signatures that were done for JSON in the now concluded JOSE WG but this time in CBOR. CBOR has new capabilities that are not present in JSON, so the intention was not to do a direct mapping from JSON to CBOR. For example as RFC 8152 mentions, COSE uses binary encoding for binary data rather than base64url, it also redefines the message structure for better identification and consistency. The group finalized the work with RFC 8152 being the target set.

**LAKE** — The Lightweight Authenticated Key Exchange (LAKE) WG was chartered at the end of 2019, with the sole purpose of designing the requirements and a LAKE solution for OSCORE that can be suitable for constrained environment as well. Once this milestone has been achieved, the WG will be closed. The LAKE WG must necessarily coordinate with other IoT-related IETF WGs such as ACE, CORE, 6TISCH, LPWAN, and LWIG in order understand and validate the requirements and the solution.

**RATS** — The Remote ATtestation ProcedureS (RATS) WG has recently been chartered in order to create an architecture, protocols and data models that allow various parties to asses the trustworthiness of remote system components. The group has already produced a draft architecture document and compressed token with signed claims about an -IoT- device.

The IETF standardization efforts in the IoT rely on the activities of additional WGs and other IETF-related organizations. In the next section, we shed light on other IETF WGs and RGs whose activities, by extension, consider the IoT landscape as well. Then we introduce how additional IETF committees and organizations are promoting network and infrastructure convergence in the IoT domain.

## V. EXPERIMENTAL, USE CASES, AND INFRASTRUCTURES

In the experimental and use cases domains, there are at least three WGs whose interests are closely linked to IoT.

**IPWAVE** — In the context of modern vehicular systems, the IP Wireless Access in Vehicular Environments (IPWAVE) WG (established in 2016) concentrates its focus on Vehicle-to-Vehicle Communications (V2V) and Vehicle-to-Infrastructure (V2I) use-cases. The main scope of the WG is to develop an IPv6 based solution to establish efficient and secure connectivity between vehicles and other infrastructure components. V2V and V2I communications may encompass the use of different link layer technologies including 802.11-OCB (Outside the Context of a Basic Service Set), 802.15.4 with 6LoWPAN, LTE-D, LPWAN, and so on. Many of these link layers already provide full support for IPv6 unlike 802.11-OCB, despite representing one of the most promising link layer in this context. In this regard, IPWAVE aims to actively coordinate with IEEE 802.11 for fully enabling transmission of IPv6 datagrams over IEEE 802.11-OCB mode. The WG also aims to document state of the art and use cases of IPv6-based vehicular networks by taking also into account the IoT protocol suite being developed in other IETF and IRTF groups. Table IV summarizes a relevant I-D defined in the IPWAVE WG.

**HOMENET** — The Home Networking (HOMENET) WG mainly focuses on automatic configuration of IPv6 within and among home networks, as well as investigating home network setup issues caused by the use of different networking technologies and heterogeneous devices. Two relevant aspects are considered. The first relates to the complexity of managing multiple home network segments, including scenarios in which the subnets are physically deployed in different places. These issues are generated, for example, by the concurrent use of different link layers (e.g., Ethernet technology and others designed for low-powered sensors) that require different routing and security policies. The second aspect regards the IPv6 support in home network devices (e.g., gateways and routers), as the use of IPv6 implies the re-definition of specific requirements and mechanisms (e.g., prefix configuration for routers, managing routing, name resolution, service discovery, network security). HOMENET has already produced several informational and proposed standard RFCs. Table IV summarizes two of the most relevant ones (RFC 7368 and RFC 7788).

**LWIG** — The main scope of the Lightweight Implementation Guidance (LWIG) WG is to provide guidelines for efficient implementation techniques and practical considerations about the usage of Internet protocols in low-capabilities devices operating in constrained environments. Defining this type of good practices becomes crucial if we consider the limited computing power, battery capacity, available memory, or communications bandwidth of constrained (IP-capable) devices. In fact, a minimal optimization in the software implementation and execution (even in the order of few kilobytes of code) can result in tangible benefits in terms of reduced complexity, memory footprint, and power usage. The WG was created with the purpose of scrutinizing IoT-related IETF protocols, such as IPv6, 6LoWPAN, CoAP, and RPL. LWIG seeks to provide recommendations for the optimization of already existing and stable protocols implementations, ensuring the critical requirement of interoperability with other devices. Table IV summarizes an RFC defining common terminology for constrained-node networks, as well as drafts outlining guidance for CoAP implementation or TCP usage guidance in IoT contexts.

Two additional IRTF groups show interest on exploring the applicability of specific network infrastructure also to IoT scenarios: ICNRG (Information-Centric Networking RG) and DINRG (Decentralized Internet Infrastructure RG).

**ICNRG** — ICNRG considers also the IoT, as domain in which to take advantage of ICN. RFC 7476 outlines what ICN characteristics can be exploited in IoT (e.g., naming, data discovery, caching), while RFC 7927 covers the research challenges deriving from such integration. Both RFCs provide a comprehensive literature review and analysis of ongoing research efforts that are aiming to enable the ICN-based IoT paradigm. It is also emphasized how IoT exposes ICN abstractions to an even stricter set of requirements due to the large quantity of nodes, as well as type and volume of information to be processed. Table IV mentions also two I-Ds, also mentions an I-D characterized by a clear IoT focus, as it specifies dedicated mechanisms for adapting ICN to LoWPAN networks.

TABLE IV: Selection of relevant RFCs and I-Ds in the experimental and use cases and infrastructures.

| RFC number and Title | Description |
|---|---|
| *ipwave* | |
| RFC 8691 *Basic Support for IPv6 Networks Operating Outside the Context of a Basic Service Set over IEEE Std 802.11* | Describes the definition of key parameters for enabling an efficient transmission of IPv6 datagrams over IEEE 802.11-OCB networks. Among these parameters the supported Maximum Transmission Unit (MTU) size, the header format preceding the IPv6 header, etc. |
| *homenet* | |
| RFC 7368 *IPv6 Home Networking Architecture Principles* | Describes principles, considerations, requirements and elements of a general architecture for IPv6-based home networking. Suggests how standard IPv6 mechanisms and addressing can be employed in home networking, together with the protocol extensions that are needed for ensuring additional functionality in residential home networks. |
| RFC 7788 *Home Networking Control Protocol* | Defines requirements for home network devices and the Home Networking Control Protocol (HNCP). HNCP provides multiple functionality such as e.g. automated configuration of addresses, name resolution, service discovery etc. |
| *lwig* | |
| RFC 7228 *Terminology for Constrained-Node Networks* | Defines the terminology for constrained-node networks, which is commonly used in standardization activities. |
| RFC 8352 *Energy-Efficient Features of Internet of Things Protocols* | Outlines the main challenges for ensuring energy-efficient protocol operations on constrained devices, together with a set of guidelines that allow overcoming such challenges. The document mainly focuses on link-layer, however, it also provides an overview of energy-efficient mechanisms available at each layer of the IETF protocol suite specified for constrained-node networks. |
| RFC 8387 *Practical Considerations and Implementation Experiences in Securing Smart Object Networks* | Provides an overview on the challenges that are faced for securing resource-constrained smart object devices and trade-offs generated by the use of certain security approaches. It also recommends a deployment model in which devices can signing message objects, querying about availability of suitable cryptographic libraries, and providing usage experience of those libraries. |
| **Draft Title** | **Description** |
| *ipwave* | |
| *IPv6 Wireless Access in Vehicular Environments (IPWAVE): Problem Statement and Use Cases* | Details problem statement and use cases on IP-based vehicular networks, considering several communications scenarios – vehicle-to-vehicle (V2V), vehicle-to-infrastructure (V2I), and vehicle-to-everything (V2X). It also analyzes proposed protocols for IP-based vehicular networking, with a particular focus on key aspects such as IPv6 Neighbor Discovery, Mobility Management, and Security & Privacy. |
| *lwig* | |
| *Building Power-Efficient CoAP Devices for Cellular Networks* | Provides guidelines for the use of CoAP in the context of smart building and scenarios in which sensors communicate through cellular networks, with a special focus on optimized techniques for minimizing the power consumption. |
| *TCP Usage Guidance in the Internet of Things (IoT)* | Provides guidelines for lightweight implementation and use of TCP in Constrained Networks. It explains several techniques to simplify a TCP stack, discussing also drawbacks. |
| **RFC number and Title** | **Description** |
| *icnrg* | |
| RFC 7476 *Information-Centric Networking: Baseline Scenarios* | Defines a set of scenarios that can be used as a base for the evaluation of different ICN approaches. It discusses also how ICN solutions can help satisfying general network requirements. |
| RFC 7927 *Information-Centric Networking (ICN) Research Challenges* | Focuses on the description of current research challenges in the ICN domain, including naming, security, routing, system scalability, mobility management, network management etc. ICN challenges in the specific context of IoT are also discussed. |
| **Draft Title** | **Description** |
| *icnrg* | |
| *ICN Adaptation to LoWPAN Networks (ICN LoWPAN)* | Defines defines a convergence layer for CCNx and NDN over IEEE 802.15.4 LoWPAN networks. The document defines stateless and stateful compression schemes to improve efficiency on constrained links. |

**DINRG** — DINRG is one of the latest RGs established in IRTF. The group's charter identifies the WG scope in investigating infrastructure services that benefit from decentralization or are difficult to realize in connectivity-constrained networks. Both aspects can relate to IoT, since centralized deployments can represent a limitation to many IoT use cases as well as well-known connectivity limitation of IoT networks. Infrastructure services decentralization brings additional research challenges (e.g., trust management, name resolution, resource discovery), which are investigated by DINRG.

## VI. SUPPORTING ORGANIZATIONS

**IoT Directorate** — Beyond the activities of the WGs focusing on IoT scenarios, the entire protocol stack is evolving and many of the new technologies developed in other IETF WGs can find applicability also in IoT contexts. With a view of strengthening an effective coordination between WGs and external IoT standard organizations and alliances, IETF has established the IoT Directorate, an advisory group of technical experts the primary purpose of which is *"coordination within IETF on IoT-related work and increasing the visibility of IETF IoT standards visibility to other SDOs, industry alliances, and other organization"* [6].

**Internet Architecture Board (IAB)** — The IAB is a committee established to serve and support the IETF activities, in order to facilitate the standardization process. In the specific context of IoT, IAB has organized multiple workshops (e.g., about security, architecture, and semantic interoperabil-ity [13]), with the intention of promoting IoT-related IETF activities and receiving active engagement of external parties for the development of a better IoT. Furthermore, it has also drawn up an informational document that offers guidance for designing Internet-connected smart objects networks [14].

**Internet Assigned Numbers Authority (IANA)** — It is an organization that coordinates and manages domain names (e.g., through DNS Root), IP addressing, as well as maintaining the Internet protocols' numbering systems where, for example, protocols' parameters specified in RFCs are defined. IANA is responsible of assigning and registering the static fields defined in the different protocols. In the case of IoT, several protocol constants require a well-kept registry (e.g., CoAP Content-Formats, COSE Algorithms, 6LoWPAN Capability Bits, and CBOR Tags), which are regularly defined during the publication of new RFCs [15].

## VII. CONCLUSION

IETF is a leading entity for specification and documentation of key IoT standards, as well as guidelines for their implementations. The IETF focus on IoT is constantly growing, besides strengthened by a larger adoption of its IoT protocol suite across other SDOs, organizations, and consortia. The main goal of this article was to provide a comprehensive and detailed overview on the current IETF activities in support of IoT deployments. We have inspected six meta-domains, presenting what is the main focus of distinct working (and research) groups operating in each single area. We have also presented a concise summary of the main outcomes produced by the WGs, as well as some of the most significant technical areas currently explored.

**Roberto Morabito** is Experienced Researcher at Ericsson Research in Finland and Research Associate with Princeton University. Roberto's interests include design and development of research projects in the context of Internet of Things, Edge Computing, and Distributed Artificial Intelligence. He holds a Doctorate degree in Networking Technology awarded by Aalto University.

**Jaime Jiménez** is Master Researcher at Ericsson and co-chair of the IETF CoRE Working Group. He has been also chair and key contributor of the IPSO Semantic Group, as well as one of the authors of the OMA-LWM2M specification. His current research interests include IoT Device and Network Management.